\documentstyle[epsfig]{aa}

\begin{document}
 
\thesaurus{12(12.03.1; 12.03.4)}

\title{Models of Universe with an inhomogeneous Big Bang singularity}
\subtitle{II. CMBR 
dipole anisotropy as a byproduct of a conic Big-Bang singularity}

\author{Jean Schneider 
\and Marie-No\"elle C\'el\'erier}

\institute{CNRS UPR 176 - D\'epartement d'Astrophysique Relativiste et de 
Cosmologie, Observatoire de Paris-Meudon, \\
5 place Jules Janssen, 92195 Meudon C\'edex, France}
 
\offprints{M.-N. C\'el\'erier}

\mail{celerier@obspm.fr}
%% cette commande marche avec la macro aa et non l-aa

\date{Received 1998 / Accepted}

\maketitle
\markboth{J. Schneider \&\ M.N. C\'el\'erier: Models of 
Universe with an inhomogeneous Big-Bang singularity}
{J. Schneider \&\ M.N. C\'el\'erier: Models of 
Universe with an inhomogeneous Big-Bang singularity}

\begin{abstract} 
The existence of stars and galaxies reqires cosmological 
models with an inhomogeneous matter and radiation distribution. But in 
these models the initial singularity surface  $t_0(r)$ is in general 
homogeneous (independent
of $r$). In this second paper of a series devoted to an inhomogeneous Big Bang 
singularity, we investigate the cosmic microwave background
radiation (CMBR) dipole. 
A special Tolman-Bondi Universe is used to study the
effect of a Big-Bang singularity, depending linearly on $r$, on the CMBR 
anisotropy. It is shown that, for an observer located off the ``center'' of 
this Universe ($r=0$), the parameters of the model can be tuned so as to
reproduce, with a good approximation, the dipole and the quadrupole moments
of the CMBR anisotropy observed in recent experiments. If the dipole should
prove cosmological, a slight delaying of the Big-Bang over spatial
coordinates would thus be a good candidate for its interpretation.

\keywords{cosmic microwave background - cosmology: theory}
\end{abstract}

\section{Introduction}

The standard cosmological models rest mainly on a homogeneous and
spherically symmetric Roberston-Walker metric $f_{\mu \nu }$, subject to the 
Einstein equations,
with a matter-energy tensor depending only on the parameter (called the cosmic 
time \footnote{Although this is an inappropriate denomination 
(Schneider, 1994)}) labelling the 3-surfaces. \\

The particular choice of one model lies in the choice of  an equation of state.
Inflationary cosmologies are part of this framework with peculiar choices of
the equation of state imported from particle physics. Departures from this
standard framework are developed for instance by the use of generalizations of
the Einstein equations or by ta\-king into account inhomogeneities in the
matter distribution. Whereas there are presently no 
compelling reasons to abandon the
Einstein equations, the introduction of inhomogeneities is unavoidable since
there are large and small-scale structures in the observed Universe. These
inhomogeneities induce inhomogeneities in the metric. They are used for the
study of the formation of galaxies and large scale structures and of the
fluctuations of the cosmic background at 3$K$ (CMBR). In these models the 
metric $g_{\mu \nu }$ is a weakly perturbed Robertson-Walker metric:
$g_{\mu \nu }=f_{\mu \nu }+h_{\mu \nu }$. In this framework the perturbation 
$h_{\mu \nu }$ is generally small and depends on the spatial coordinates, 
but the main term $f_{\mu \nu }$, characterized by a scale para\-meter $R(t)$ 
is independent of position. Therefore the cosmological singularity $R=0$ is 
an equal time 3-surface; in astronomical terms, the age of the Universe is 
the same everywhere.  \\

Solutions of the Einstein equations with inhomogeneous $R=0$ 
hyper-surfaces have been studied analytically by Tolman (1934) and Bondi 
(1947).
The Tolman-Bondi universes have applied to cosmological contexts such as 
clusters of galaxies (Tarentola, 1976) or the CMBR dipole (Paczynski and Piran,
1990). \\

But the parameters of the model are then set in such a way that they lead to a
spatially homogeneous $R=0$ Big Bang singular surface.  There are no more 
reasons for this choice than for a strictly homogenous matter and radiation 
distribution. For instance, if the present universe were 
the result of the bounce from a collapse prior to the standard 
Big Bang (a solution which cannot be excluded), a strictly homogeneous $R=0$ 
surface would result from a very unlikely fine tuning. \\

In a recent work (C\'el\'erier \& Schneider, 1998), we have identified a class 
of inhomogeneous models of Universe, with a Big-Bang of ``delayed'' type, 
solving the standard horizon problem without need for an inflationary phase. 
In the present paper, we investigate the application of a peculiar model of 
this class to the CMBR anisotropy. \\

From a purely geometrical point of view, it is always possible to re-label the 
3-surfaces so as to make the hyper-surface $R=0$ independent of $r$. But
such an arbitrary re-labelling is forbidden by the description of
physical phenomena by the Schr\"odinger equation in curved space time.
This equation, which gives the rate of evolution of phenomena, in particular, 
the thermal history of the Universe (through nucleosynthesis of light 
elements and matter-radiation decoupling), provides, in an inhomogeneous 
Universe, a clock imposing on time-like
world lines a given time coordinate. Each hamiltonian used in the 
Schr\"odinger equation gives a different time scale ({\it e.g.} the atomic
transition rates) which depends on the local curvature. But there is an 
implicit postulate, that there is a fundamental time scale, the Planck time 
$\sqrt{Gh/c^3}$. Thus, the time coordinate can only be, with the
choice of a given time unit, rescaled globally with a universal affine
transformation $t\longrightarrow at+b$ where the coefficient $a$ and $b$ are 
independent on $r$ and $t$. \\

The dipole moment in the CMBR anisotropy is the most prominent feature in 
the recent observational data, as probed by the four years COBE experiments
\footnote{The three experiments aboard the COsmic Background
Explorer satellite (COBE) are the Far-InfraRed Absolute Spectrophotometer
(FIRAS) 60 - 630 GHz, the Differential Microwave Radiometers (DMR)
30 - 90 GHz and the Diffuse InfraRed Background Experiment (DIRBE)
1.2 - 240 $\mu m$. All experiments provide maps of small temperature
fluctuations from an average 2.73$^\circ$ K for the CMBR.}. It overcomes the
quadrupole, of order 5.10$^{-6}$, by more than two orders of magnitude,
its value being of order 10$^{-3}$ (Smoot et al., 1992; Kogut et al., 1993).\\

This dipole is usually considered as resulting from a Doppler effect
produced by our motion with respect to the CMBR rest-frame (Partridge,
1988). A few authors (Gunn, 1988; Paczynski \&\  Piran, 1990; Turner, 1991;
Langlois \&\  Piran, 1996; Langlois, 1996), in the recent past, intended
however to show that its origin could be in the large scale features of the 
Universe.\\

Paczynski and Piran (1990), using an ad hoc toy model, emphasized the
possibility for the dipole to be ge\-nerated
by an entropy gradient in a Tolman-Bondi dust Universe. In the peculiar
model they did study, they have assumed that the time of the Big-Bang was the
same for all observers. \\

Hereafter we show that the dipole, and quadrupole,
anisotropy, or part of it, could be considered as the outcome of a conic
Big-Bang surface. \\

We first describe, in the following section, the special Tolman-Bondi
model we use for our derivation. The calculations will be developed in
Sect.3 and the results exposed in Sect.4. Our conclusions and a brief
discussion are given in Sect. 5.

\section{A flat dust spherically symmetrical model}

We consider here the light cone emitted from the last scattering
surface - temperature of order 4.10$^3$ K - towards the Earth at
our present time. Since this period is matter dominated - the radiation
was dynamically relevant only at times prior to a temperature of
order  10$^4$ K - we are considering the behaviour of a photon gas
immersed into an Universe satisfying \\
$$\rho_{\rm radiation} << \rho_{\rm dust}$$ \\
that is, we neglect the radiation as source of gravitational field.\\

We then choose a Tolman-Bondi (Tolman, 1934; Bondi, 1947) model which
figures out a dust (ideal non zero rest mass pressureless gas)
dominated, spatially spherically symmetrical inhomogeneous Universe.\\

The Bondi line element, in co-moving coordinates and proper time, is:
\begin{equation}
ds^2=-c^2dt^2+S^2(r,t)dr^2 + R^2(r,t)(d\theta ^2+\sin^2 \theta d\varphi^2)
\label{1}
\end{equation}

It reduces to the usual Friedmann-Robertson-Walker metric for a
homogeneous Universe.\\

Solving Einstein equations for this metric gives:
\begin{equation}
S^2(r,t) = {R^{'2}(r,t) \over 1+2E(r)/c^2} \label{2}
\end{equation}
\begin{equation}
{1\over 2} \dot{R}^2 (r,t) - {G M(r)\over R(r,t)} = E(r) \label{3}
\end{equation}
\begin{equation}
4\pi \rho (r,t) = {M'(r)\over R'(r,t)R^2(r,t)} \label{4}
\end{equation}
a dot denoting differentiation with respect to $t$ and a prime
differentiation with respect to $r$.\\

$\rho (r,t)$ is the energy density $\rho_{\rm dust}$\\

$E(r)$ and $M(r)$ are arbitrary functions of $r$. $E(r)$ can be interpreted
as the total energy per unit mass and $M(r)$ as the baryonic mass within
the sphere of co-moving radial coordinate $r$. $M(r)$ remaining constant
with time, we use it to define the radial coordinate: $M(r) \equiv M_0r^3$,
where $M_0$ is a constant.\\

Eq.(3) can be solved and gives a parametric expression for $R(r,t)$ in
case $E(r) \not= 0$ and an analytic one in case $E(r)=0$. \\

We retain the flat Universe model $E(r)=0$ and the analytic expression: \\
$R(r,t)=[9G M(r)/2]^{1/3}[t-t_0(r)]^{2/3}$ \\
which, with the above definition for the radial coordinate, becomes:
\begin{equation}
R(r,t)=(9GM_0/2)^{1/3}r[t-t_0(r)]^{2/3}  \label{5}
\end{equation}

The homogeneous limit of our model is the Einstein-de Sitter Universe
with $\Omega =1$.\\

$t_0(r)$ is another arbitrary function of $r$. It is the Big-Bang
hyper-surface.\\

A class of models, identified as solving the horizon problem (C\'el\'erier \& 
Schneider, 1998), exhibits a Big-Bang function of the form:
\begin{equation}
t_0(r) = br^n  \mbox{}\hspace{1.cm}b>0, n>0
\end{equation}

We here choose, for simplicity, to investigate the pro\-perties of the 
subclass:

\begin{equation}
t_0(r) = br \mbox{}\hspace{0.5cm} {\rm with} \mbox{}\hspace{.5cm}1/R_H>b>0
\end{equation}
This conic surface corresponds to perturbations with low ($k<1/R_H$, $R_H$ 
being the horizon radius) spatial frequencies.

Eq. (5) thus becomes:
\begin{equation}
R(r,t)=(9GM_0/2)^{1/3} r(t-br)^{2/3} \label{7}
\end{equation}

The specific entropy $S$ is usually defined as the ratio of the number
density of photons over the number density of baryons:
\begin{equation}
S\equiv {k_Bn_{\gamma}(r,t) m_b\over \rho (r,t)} \label{8}
\end{equation}
where $m_b$ is the baryon mass and $k_B$, the Boltzmann cons\-tant.

In order to decouple the effect of inhomogeneous entropy distribution (as
already studied by Paczynski and Piran, 1990) from an inhomogeneous $R=0$ 
surface, we retain:\\
$S=\hbox{const}$. \\

The observed deviation of the CMBR from a perfect homogeneous pattern being
very small, we can assume, as a reasonable approximation, thermodynamical
equilibrium for the photons, so as to write, at the ultra-relativistic
limit for bosons:
\begin{equation}
n_\gamma = a_n T^3  \label{9}
\end{equation}
$T$ being the radiation temperature and 
\mbox{$a_n=2\zeta(3)k_B^3/[\pi ^2(\hbar c)^3]$.}\\

Letting, with no loss of generality, $S=\hbox{const.}=k_B\eta _0$, and taking
the present
photon to baryon density ratio $\eta _0$ to be $10^8/(2.66\Omega _bh_0^2)$, 
we derive
the following expression for $T$:
\begin{equation}
T(r,t)=\left( \frac{10^8}{2.66h_0^22\pi Ga_nm_b (3t-5br)(t-br)}\right)^{1/3}
\label{10}
\end{equation}
where $h_0$ is the Hubble constant in units 100 km/s/Mpc. Hereafter, for 
numerical applications, we take $h_0$ = 0.75.
\medskip

\section{Integration of the null geodesics and determination of the dipole
and quadrupole moments}

Let an observer, for example the COBE satellite, be located at $(t_0,r_0)$
where the average temperature is $T_0$, of order 2.7 $K$. The
radial co-moving coordinate $r_0$ is choosen to be non zero so as to
put the observer off the center of the Universe.\\

The light travelling from the last scattering surface to this observer follows
null geodesics which we are going to numerically integrate, from the 
observer, until we reach this 
surface defined by its temperature $T_{\ell s}=4.10^3 K$. \\

In principle, one should integrate the optical depth equation along with the 
null geodesic equations. Here, we approximate the optical depth by a 
step-function. This procedure leads to integrate the null geodesics until the 
temperature reachs $T_{\ell s}=4.10^3 K$. \\

Our toy Universe being spherically symmetrical, an observer at a distance
from the center sees an axially symmetrical Universe in the center
direction. It is thus legi\-timate to integrate the geodesics in the
meridional plane. The photons path is uniquely defined
by the observer position $(r_0,t_0)$ and the angle $\alpha$ between
the direction from which comes the light ray as seen by the observer and
the direction towards the center of the Universe.\\

In the following, we adopt the units:\\
$c=1$, $8\pi G/3=1$ and $M_0=1$.\\

For the metric given by Eq.(1), the meridional plane is defined as:\\
$\theta =\pi /2 \qquad \sin\theta =1 \qquad k^\theta =0$\\
$k^\theta$ being the $\theta$ component of the photon wave-vector defined as:\\
$k^\mu =-{dx^\mu \over d\lambda}$\\
which gives:
\begin{equation}
{dt\over d\lambda}=-k^t \label{11}  \\
\end{equation}
\begin{equation}
{dr\over d\lambda}={k_r\over R'^2}=(16\pi /27)^{2/3} ~{9(t-br)^{2/3}\over
(3t-5br)^2} k_r \label{12}                \\
\end{equation}
\begin{equation}
{d\varphi\over d\lambda}={k_\varphi\over R^2} \label{13}
\end{equation}

From the geodesic equations of light:\\
$dk^\mu /d\lambda + \Gamma_{\nu \lambda}^\mu k^\nu k^\lambda =0$\\
we obtain after some calculations:
\begin{eqnarray}
{dk^t\over d\lambda}&=&-2\left({16\pi \over 27}\right)^{2/3}
\left[{3(3t-2br)\over (3t-5br)^3(t-br)^{1/3}}(k_r)^2 \right. \nonumber \\
&+&\left.  {1\over 3r^2(t-br)^{7/3}} (k_\varphi)^2\right]
\label{14}\\
{dk_r \over d\lambda}&=&-\left({16\pi \over 27}\right)^{2/3}
\left[{6b(2t-5br)\over (3t-5br)^3(t-br)^{1/3}} (k_r)^2 \right. \nonumber \\
 &+& \left. {3t-5br \over 3r^3(t-br)^{7/3}} (k_\varphi)^2 \right]
\label{15}
\end{eqnarray}
\begin{equation}
k_\varphi=\hbox{const.}     \label{16}
\end{equation}

For photons: $ds^2=0$ coupled with Eq.(12) to (14) gives:
\begin{equation}
(k^t)^2=\left({k_r \over R'}\right)^2 + \left({k_{\varphi}\over R}\right)^2
\label{17}
\end{equation}

The equation for the redshift $z_{\ell s}$ in co-moving coordinates is:\\
\[
1+z_{\ell s}={(k^t)_{\ell s} \over (k^t)_0}
\]
$(k^t)_{\ell s}$ and $(k^t)_0$ being the time-like component of the
photons wave-vector at the last-scattering and at the observer respectively.

The former equations system can be integrated, the following initial
conditions being given at the observer:
\begin{equation}
t=t_0 \qquad r=r_0 \qquad (k^t)_0=1  \label{18}
\end{equation}

And thus:
\begin{equation}
1+z_{ls}=(k^t)_{ls} \label{19}
\end{equation}

At a given couple $(t_0,r_0)$ corresponds values for $R$ and its partial
derivatives at $t_0,r_0$.\\

We denote:
\begin{eqnarray}
R_0&\equiv & R(t_0,r_0) \quad \hbox{given by Eq.(8)} \nonumber \\
R'_0&\equiv &R'(t_0,r_0) \quad \hbox{and so on} \nonumber
\end{eqnarray}

The observer at $(t_0,r_0)$ seeing the photons trajectory making an angle
$\alpha$ with the direction towards the center of the Universe, we can
write:
\[
(k_r)_0=A \cos \alpha \qquad (k_\varphi)_0=B \sin \alpha \nonumber
\]

Substituting the former values of the coordinates of ${\bf k_0}$ into Eq.(18)
written at $(t_0,r_0)$, we find:
\[
A=R'_0 \qquad B=R_0  \nonumber
\]

And thus:
\begin{equation}
(k_r)_0=R'_0 \cos \alpha \qquad (k_\varphi)_0=R_0 \sin \alpha \label{20}
\end{equation}

Eq.(17) becomes:
\[
k_\varphi= R_0 \sin \alpha \nonumber
\]

Substituting in Eq.(18), we get:
\[
(k_r)^2 = R'^2 [(k^t)^2-(R_0 \sin \alpha / R)^2]  \nonumber
\]
which possesses two solutions:
\begin{equation}
k_r=\pm R'[(k^t)^2-(R_0 \sin \alpha / R)^2]^{1/2} \label{20}
\end{equation}

From Eq.(4), with $M(r)\equiv r^3$, comes:
\[
\rho_{\rm dust} = (3/4\pi) {r^2\over R'R^2} \nonumber
\]

As we want, for physical consistency, $\rho_{\rm dust} \geq 0$, we get:
\[
R' \geq 0 \nonumber
\]

And because Eq.(13) implies the same sign for $dr/d\lambda$ and $k_r$,
it follows that:\\

Eq.(22) with the plus sign is the solution $dr/d\lambda >0$, where $r$
is increasing with increasing $\lambda$ parameter.\\

Eq.(22) with the minus sign is the solution $dr/d\lambda <0$, where
$r$ is decreasing with increasing $\lambda$.\\

Substituting Eq.(22) into Eq.(12) to (17), we get, after some calculations,
the reduced system of three differential equations:
\begin{equation}
{dt\over d\lambda}=-k^t \label{22}
\end{equation}
\begin{eqnarray}
{dr\over d\lambda}&=&\pm \left({16\pi \over 27}\right)^{1/3}
{3(t-br)^{1/3}\over 3t-5br} \nonumber \\
&&\left[(k^t)^2 -{r^2_0(t_0-br_0)^{4/3}\sin ^2 \alpha
\over r^2(t-br)^{4/3}} \right]^{1/2} \label{23}
\end{eqnarray}
\begin{equation}
{dk^t\over d\lambda}={2(3t-2br) \over 9(t-br)(3t-5br)}(k^t)^2 +
{2br^2_0(t-br_0)^{4/3}\sin ^2 \alpha \over r(3t-5br)(t-br)^{7/3}}
\label{24}
\end{equation}

Provided we choose the affine parameter $\lambda$ increasing from $\lambda=0$
at $(t_0,r_0)$ to $\lambda = \lambda_{\ell s}$ at $(t_{\ell s}, r_{\ell s})$ on
the last scattering surface, we have to consider two cases:\\

- the ``out-case": the observer looks at a direction opposite to the center
of the Universe $(\alpha > \pi /2)$. We thus integrate the null geodesics
from $(t_0,r_0)$ to $(t_{\ell s},r_{\ell s})$ with an always increasing $r$. We
have to retain the plus sign in Eq.(24).\\

- the ``in-case": the observer looks at a light ray first approaching the
center of the Universe, then moving away from it before reaching her eyes
$(\alpha <\pi/2)$. Eq.(24) with the minus sign first obtains until
$dr/d\lambda =0$, then the minus sign in Eq.(24) changes to a plus sign. \\

As, in the system of Eq.(23) to (25), the dependence in $\alpha$ is of the form
$\sin \alpha$ and as $\sin \alpha = \sin (\pi - \alpha)$, we can only
discriminate between the ``out" and ``in" cases by the behaviour of the
sign of $dr/d\lambda$.\\

We integrate a number of ``in" and ``out" null geodesics, each caracterized
by a value for $\alpha$ between zero and $\pi /2$, back in time from the
observer at $(t_0,r_0,T_0)$ until the temperature, as given by Eq.(11),
reachs \\$T_{\ell s}=(4/2.7) 10^3 T_0$, which approximately defines the last
scattering surface.\\

At this temperature, the redshift with respect to the observer, as given by
Eq.(20), is $z_{\ell s}^{\rm in-out}(\alpha)$, somewhat varying, with
the $\alpha$ angle and the ``in" and ``out" direction, about an average
$z^{av}_{\ell s}$.\\

The apparent temperature of the CMBR mesured in the $\alpha$ in-out direction
is:
\[
T^{\rm in-out}_{\rm CMBR}(\alpha)={T_{\ell s}\over 1+z^{\rm in-out}_{\ell s}
(\alpha)}=T^{av}_{\rm CMBR}{1+z^{av}_{\ell s}\over 1+z^{\rm in-out}_{\ell s}
(\alpha)} \nonumber
\]
where the averages for $T$ and $z$ are calculated over the whole sky. We
write with simplified notations:
\begin{equation}
{T_{\rm CMBR}\over T^{av}}={1+z^{av}\over 1+z_{\ell s}} \label{25}
\end{equation}

The CMBR temperature large scale inhomogeneities are expanded in spherical 
harmonics:
\[
{T_{\rm CMBR}(\alpha,\varphi) \over T^{av}}= \sum^{\infty}_{\ell=1}
\sum^{+\ell}_{m=-\ell} a_{\ell m} Y_{\ell m} (\alpha,\varphi) \nonumber
\]
$\alpha$ being the Euler angle usually called $\theta$ in spherical
coordinates, and with:
\begin{equation}
a_{\ell m}=\int {T_{\rm CMBR}(\alpha,\varphi) \over T^{av}} Y^*_{\ell m}
(\alpha,\varphi) \sin \alpha ~d\alpha ~d\varphi\label{26}
\end{equation}

The dipole and quadrupole moments are defined as:
\[
D=(|a_{1-1}|^2+|a_{10}|^2+|a_{11}|^2)^{1/2} \nonumber
\]
\[
Q=(|a_{2-2}|^2+|a_{2-1}|^2+|a_{20}|^2+|a_{21}|^2+|a_{22}|^2)^{1/2}
\nonumber
\]
 
In the special case we are interested in, the large scale inhomogeneities 
only depend on the $\alpha$ angle so that all the $a_{\ell m}$ with 
$m\not= 0$ are zero.\\

The dipole and quadrupole moments thus reduce to:
\[
D=a_{10} \qquad Q=a_{20} \nonumber
\]
$a_{10}$ and $a_{20}$ being given by Eq.(27) with:
\[
Y_{10}(\alpha)=\sqrt{{3\over 4\pi}}\cos \alpha \qquad
Y_{20}(\alpha)=\sqrt{{5\over 4\pi}} \left({3\over 2}\cos ^2 \alpha -{1\over 2}
\right)      \nonumber
\]

It follows:
\begin{equation}
D=(1+z^{av}) \int^\pi_0 {Y_{10}(\alpha)\over 1+z_{\ell s} (\alpha)}
\sin \alpha ~d\alpha  \label{27}
\end{equation}
\begin{equation}
Q=(1+z^{av}) \int^\pi_0 {Y_{20}(\alpha) \over 1+z_{\ell s}(\alpha)}
\sin \alpha ~d\alpha \label{28}
\end{equation}

Taking into account Eq.(20) and the spherical symmetry of the model, we
obtain:
\begin{equation}
D=\left|{1\over 2}\sqrt{{3\over \pi}} k^t_{av} \left[\int^{\pi\over 2}
_0 {\cos \alpha \sin \alpha \over k^t_{\rm in}(\alpha)} d\alpha -
\int^{\pi\over 2}_0 {\cos \alpha \sin \alpha \over k^t_{\rm out}
(\alpha)} d\alpha \right] \right|
\label{29}
\end{equation}
\begin{eqnarray}
Q&=&{1\over 4} \sqrt{{5\over \pi}} k^t_{av} \left[\int^{\pi\over 2}_0
{(3\cos^2 \alpha - 1)\sin \alpha \over k^t_{\rm in} (\alpha)} d\alpha \right. \nonumber \\
 &+& \left. \int^{\pi \over 2}_0 {(3\cos^2 \alpha - 1)\sin \alpha \over k^t_{\rm out}
(\alpha)} d\alpha  \right] \label{30}
\end{eqnarray}

\vskip2truecm
\section{Results}

We have first numerically integrated a number of ``out'' and ``in'' null 
geodesics, for various $r_0$ and $b$, and for 
values of $\alpha$ going from 0 to $\pi\over 2$, with $t_0$ corresponding to 
$T_0=2.7 K$ in Eq.(11).\\

 We have then calculated the dipole and quadrupole
moments $D$ and $Q$, according to Eqs.(30) and (31).\\

We have selected the values of the doublets leading to $D$ and $Q$
approaching the observed values \footnote{This choice will be discussed in 
section 5.} $D\sim 10^{-3}$, $Q\sim 10^{-5}$. These
results are given in Fig. 1 and 2.\\

\includegraphics[height=8cm,width=6cm,angle=-90]{dipplot2.ps}

\begin{center}
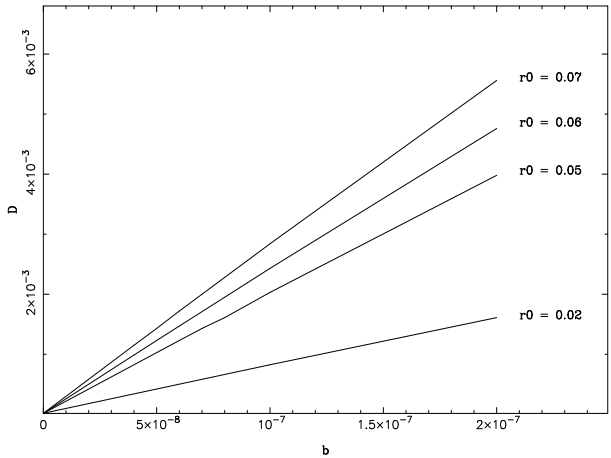
{\bf Fig. 1 : The Dipole as a function of $b$ for various $r_0$}
\end{center}

\includegraphics[height=8cm,width=6cm,angle=-90]{quadrupole2.ps}

\begin{center}
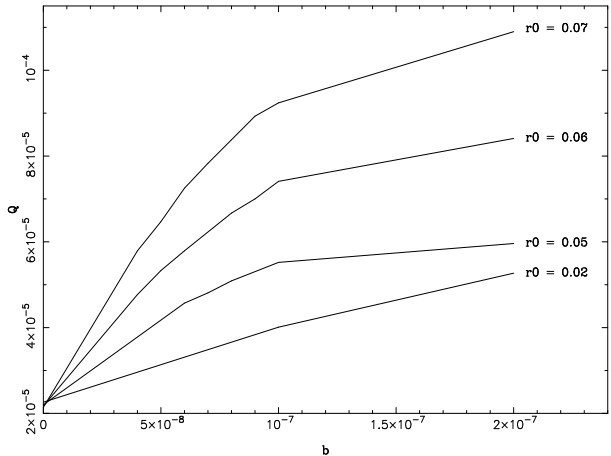
{\bf Fig. 2 : The Quadrupole as a function of $b$ for various $r_0$}
\end{center}

\bigskip

The best fitted values of $r_0$ and $b$ giving $D$ close to $10^{-3}$ and 
$Q$ close to $10^{-5}$ are shown in Table 1.

\begin{table}
\caption[]{Best fitted values of $r_0$ and $b$}
\label{table}
 \[
         \begin{array}{p{1cm}p{1.5cm}p{2cm}p{2cm}}
            \hline
            \noalign{\smallskip}
            $r_0$ & $b$ & $D$ & $Q$ \\
            \noalign{\smallskip} 
            \hline
            \noalign{\smallskip}
            0.02 & $2$ $10^{-7}$ & $1.61$ $10^{-3}$ & $5.27$ $10^{-5}$ \\
            0.03 & $9$ $10^{-8}$ & $1.11$ $10^{-3}$ & $3.70$ $10^{-5}$ \\
            0.04 & $7$ $10^{-8}$ & $1.15$ $10^{-3}$ & $3.99$ $10^{-5}$ \\
            0.05 & $6$ $10^{-8}$ & $1.23$ $10^{-3}$ & $4.57$ $10^{-5}$ \\
            0.06 & $5$ $10^{-8}$ & $1.23$ $10^{-3}$ & $5.33$ $10^{-5}$ \\
            0.07 & $4$ $10^{-8}$ & $1.15$ $10^{-3}$ & $5.79$ $10^{-5}$ \\

            \noalign{\smallskip}
            \hline
         \end{array}
     \]
   \end{table}

\vskip2truecm
\section{Conclusion and discussion}

Using a toy model, chosen within the class of delayed Big-Bang models 
identified as sol\-ving the horizon problem without need for any inflationary 
phase (C\'el\'erier \& Schneider, 1998), and presenting the following main 
features:\\
- dust dominated spherically symmetrical Tolman-Bondi Universe\\
- conic Big-Bang singularity \\
- observer located off the center of the Universe \\
we showed that can be
found values for the parameters of the model - the location of the
observer in space-time and the increasing rate of the Big-Bang function - that
allow to somehow reproduce the observed dipole and quadrupole moments
in the CMBR anisotropy.\\

This provides a new possible interpretation of the dipole (or part of it, as 
it is obvious that there is probably a Doppler component due to the
local motion of the Galaxy with respect to the CMBR rest frame).\\

As has been stressed by other authors (Paczynski \&\ Piran, 1990; Turner, 1991;
Langlois \&\ Piran, 1996; Langlois, 1996), there are various observational
ways to discriminate between a local and a cosmological origin for the
dipole.\\

From an analysis of a sparse-sampled redshift survey of IRAS Point Source 
Catalog $60- \mu m$ sources, performed with the tools of standard cosmology, 
Rowan-Robinson et al. (1990) conclued, for instance, that the peculiar 
velo\-city of the Local Group should be $579$ $\Omega_0^{0.6} km s^{-1}$ 
towards $(l,b)=(269.5,29.8)$. \\

For $\Omega_0 \sim  0.3$, this would give a velocity of order $280 km s^{-1}$, 
to be compared to the CMBR dipole velo\-city: $600 \pm 50 km s^{-1}$ 
(Partridge, 1988) towards $(l,b)=(124.7 \pm 0.8,48.2 \pm 0.5)$ (Smoot et al., 
1992). \\

In this framework, the local component of the dipole would be of order 50\% of 
the total dipole. \\

If, from future analyses of observational data, part of the dipole was 
confirmed to appear non Doppler, other work, connected in particular
with multipole moments of higher order, would be needed to discriminate
between the various cosmological candidate interpretations. \\

It has to be stressed that, if the inhomogeneous Big Bang assumption is thus 
retained, a 50\% shift in the dipole cosmological component would not 
significantly affect the results given in above Table 1. \\

In our formerly cited paper (C\'el\'erier \& Schneider, 1998), we have shown 
that the horizon problem can be solved by means of a delayed Big-Bang, 
provided 
the observer is located near the center of a spherically symetrical Universe. 
Work is in progress to extend these results to models with an observer 
arbitrarily situated off the center. \\

Another interesting feature of the here presented work is to show that, 
in a model of the above studied class, the occurrence of a cosmological 
component of the dipole implies a relation between the location $r_0$ of the 
observer and the slope $b$ of the Big-Bang function. \\

It can be seen, from Table 1, that the larger $r_0$, the smaller $b$, and this 
yields a selection within the parameters space of the conic Big-Bang models. \\

We conjecture that such a feature pertains to any subclass of the delayed 
Big-Bang models.

\end{document}